\documentclass[journal]{IEEEtran}

\usepackage{cite}
\usepackage{amsmath,amssymb,amsfonts}
\usepackage{algorithmic}
\usepackage{graphicx}
\usepackage{textcomp}
\usepackage{xcolor}

\usepackage{float}
\usepackage{mathtools}
\usepackage{algorithm}
\usepackage{amsthm}
\usepackage{tablefootnote}
\usepackage{threeparttable}
\usepackage{enumitem}
\usepackage{subcaption}

\usepackage{multirow}

\newcommand{\eqdef}{\stackrel{\triangle}{=}}
\newcommand{\be}{\begin{equation}}
\newcommand{\ee}{\end{equation}}
\newcommand{\ba}{\begin{array}}
	\newcommand{\ea}{\end{array}}

\renewcommand{\thesubsubsection}{\thesubsection.\arabic{subsubsection}}

\theoremstyle{definition}
\newtheorem{definition}{Definition}[section]
\renewcommand*{\thefootnote}{\fnsymbol{footnote}}

\ifCLASSINFOpdf
\else
\fi
\begin{document}
	
	\title {NOMA Computation Over Multi-Access
		Channels for Multimodal Sensing}
	
	\author{Michel~Kulhandjian,~\IEEEmembership{ Senior Member,~IEEE,
} Gunes~Karabulut~Kurt,~\IEEEmembership{Senior Member,~IEEE,
} Hovannes~Kulhandjian,~\IEEEmembership{Senior Member,~IEEE,
} Halim~Yanikomeroglu,~\IEEEmembership{Fellow,~IEEE,
} and Claude~D'Amours,~\IEEEmembership{Member,~IEEE
}
\thanks{M. Kulhandjian and C. D'Amours are with the School of Electrical Engineering, \& Computer Science, University of Ottawa, Ottawa, Canada, e-mail: mkk6@buffalo.edu, cdamours@uottawa.ca.}
\thanks{G. Karabulut Kurt is with the Department of Electrical Engineering,  Polytechnique Montr\'eal, Montr\'eal, Canada, e-mail: \mbox{gunes.kurt@polymtl.ca}.}
\thanks{H. Kulhandjian is with the Department of Electrical \& Computer Engineering, California State University, Fresno, U.S.A., e-mail: \mbox{hkulhandjian@csufresno.edu}.}
\thanks{H. Yanikomeroglu is with the Department of Systems \& Computer Engineering, Carleton University, Ottawa, Canada, e-mail: halim@sce.carleton.ca.}

}

	\maketitle
	
	\begin{abstract}
	An improved mean squared error (MSE) minimization solution based on eigenvector decomposition approach is conceived for wideband non-orthogonal multiple-access based computation over multi-access channel (NOMA-CoMAC) framework. This work aims at further developing NOMA-CoMAC for next-generation multimodal sensor networks, where a multimodal sensor monitors several environmental parameters such as temperature, pollution, humidity, or pressure. We demonstrate that our proposed scheme achieves an MSE value approximately $0.7$ lower at $E_b/N_o = 1$ dB in comparison to that for the average sum-channel based method. Moreover, the MSE performance gain of our proposed solution increases even more for larger values of subcarriers and sensor nodes due to the benefit of the diversity gain. This, in return, suggests that our proposed scheme is eminently suitable for multimodal sensor networks.
	\end{abstract}
	
	
	\section{{Introduction}}
	\renewcommand*{\thefootnote}{\arabic{footnote}}


Internet of Things (IoT) networks are evolving towards a wide range of applications, varying from e-health, autonomous transmission systems and smart factories with ever increasing data rate requirements and reduced latency \cite{ozgun}. Their energy efficiency also need to be high to extend the battery lifetimes as much as possible. To further complicate the design problem, the expanding number of applications introduce an ever-increasing number of devices that need to be serviced with the tight operational challenges. Unfortunately, the conventional multiple access techniques, such as time-division multiple access (TDMA), frequency  division  multiple  access (FDMA)/orthogonal FDMA (OFDMA) do not offer such scalability \cite{goldsmith}. 

A promising approach is to exploit the superposition property of the wireless multiple access channel to perform some of the functionalities associated with data collection from the sensor nodes of the IoT networks over the air, while transmitting simultaneously.  This approach, referred to as computation over multi-access channels (CoMAC), introduced in \cite{nazer}, can realize a desired function of the distributed data over the wireless channel. Its extension to practically relevant systems is later introduced in \cite{goldenbaum}, and further developed in \cite{goldenbaum2}. In CoMAC,  due to the simultaneous transmission of the sensor nodes, the transmission times are scalable. Yet due to the narrow transmission bandwidth of the classical CoMAC approach introduces limited performance improvement in terms of the  spectral efficiency. To improve the spectral efficiency and overall IoT network throughput, a wideband CoMAC is proposed in \cite{wu2020noma}, integrated with the power domain non-orthogonal multiple-access (NOMA) technique. By making use of NOMA-based computation over multi-access channel (NOMA-CoMAC) technique, the authors show that the computation rate can be improved. Yet, the optimization of NOMA-CoMAC is not considered in the literature in terms of obtaining the minimum mean square error (MMSE). It is widely known that the MMSE performance can be improved by optimizing the transmission, as shown for narrowband CoMAC in \cite{semiha}. However, a wideband-CoMAC design approach to obtain the MMSE solution has not been explored to the best of our knowledge.

To address this gap in the literature, in this work we introduce a mean squared error (MSE) minimization based optimization problem of NOMA-CoMAC. Our contributions are summarized as follows:
      \vspace{-0.0cm}
      \begin{enumerate}[label=(\arabic*)]
\item  We develop a MSE optimization criterion for the wideband NOMA-CoMAC framework.
\item We propose an eigenvector-based solution to the MSE optimization problem. 
\item We show through simulation studies that our proposed scheme achieves around $0.7$ lower at $E_b/N_o = 1$ dB in terms of MSE performance compared to average sum-channel based method.
\end{enumerate}

	The rest of the paper is organized as follows. In Section \ref{SystemModel}, we discuss the NOMA-CoMAC framework, the formulation of MSE optimization problem and our proposed solution approach. After illustrating simulation results in Section \ref{simulations}, main conclusions are drawn in Section \ref{conclusion}.
	
	The following notations are used in this paper. All boldface lower
	case letters indicate column vectors and upper case letters indicate
	matrices, $()^T$ denotes the transpose operation, $()^H$ represents the conjugate transpose operation and $\mathbb{E} \{ \cdot \}$ denotes the expected value.
%


	\begin{figure*}
		\centering
		\includegraphics[width=6.0 in]{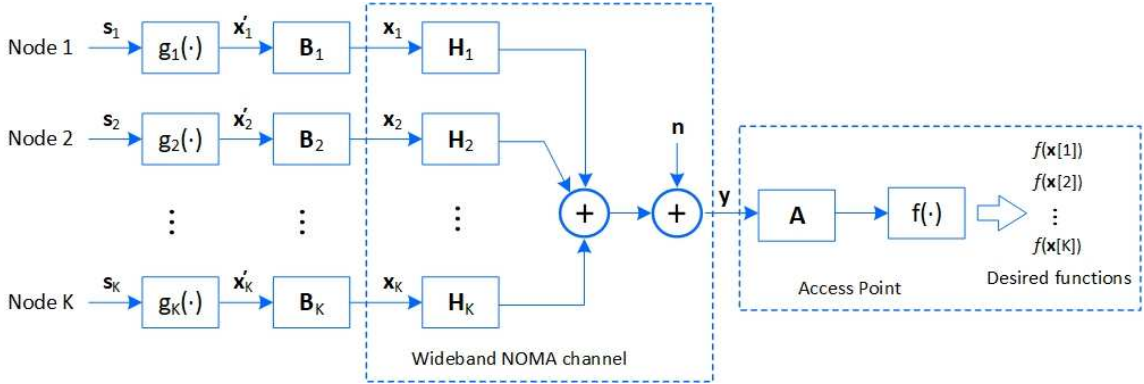}
		\centering \caption{System model CoMAC via NOMA network.} \label{CoMACSystem01}
		\vspace{-0.2cm}
	\end{figure*}
	
	\section{System model}
	\label{SystemModel}
	 In this paper, we consider a fusion technique of the multimodal sensing that aims to model context from different modalities effectively by entailing the combination of the heterogeneous sensors. The proposed multimodal sensing results in achieving improved accuracy and more specific inferences than could be achieved by the use of a single sensor alone \cite{Castanedo2013}. 
	
	Based on the multimodal sensing benefits, we study a wireless sensor network consisting of $K$ multimodal sensors and a single access point (AP).	At each node multimodal sensors record the values of $P$ heterogeneous time-varying parameters of the environment, e.g., temperature, pollution, humidity, or pressure. The measurement vector of the $k$-th sensor node constitutes $P$ sample values and is denoted by $\mathbf{s}_k = [s_{k,1}, s_{k,2}, \dots, s_{k, P}]^T \in \mathbb{R}^{P \times 1}$, where $s_{k,p}$ is the measurement of the parameter $p$ at the $k$-th sensor. Rather than accumulating the multimodal data set, the AP aims at computing $P$ functions of $P$ corresponding measuring data types from $K$ sensor nodes, denoted by $\{h_p (s_{1,p}, s_{2,p}, \dots, s_{K, p}) \}_{p=1}^P$. A class of nomograpic functions of the distributed data can be carried out quite efficiently with the aid of CoMAC. 
	\begin{definition}
	The function $h_p (s_{1,p}, s_{2,p}, \dots, s_{K, p}) $ is defined as nomographic, if there exist $K$ preprocessing functions $g_{k,p}(\cdot)$ and a postprocessing function $f_p(\cdot)$ such that it can be represented in the form \vspace{-0.2cm}
	\begin{equation}
	    h_p (s_{1,p}, s_{2,p}, \dots, s_{K, p}) = f_p\left (\sum_{k=1}^K g_{k,p}(s_{k,p})\right ).
	    \label{nomographicF}
	\end{equation}
		\end{definition}
		\begin{center}
		\vspace{-0.2cm}

\end{center}
		By exploiting the fact that wireless sensor networks normally aim to obtain a function value of sensor readings (e.g., arithmetic mean, geometric mean, etc.) instead of requiring all readings from the sensors, CoMAC framework becomes suitable for such computations. Motivated by this fact, we propose a multimodal sensor network system for future IoT networks based on CoMAC scheme over the NOMA channels, which is portrayed in Fig. \ref{CoMACSystem01}. Explicitly, the readings at each sensor nodes are preprocessed by specified functions $g_k(\cdot) =\{ g_{k,p}(\cdot)\}$, where $g_{k,p}(\cdot)$ operates on $s_{k,p}$ and $g_k(\mathbf{s}_k) = [g_{k,1}(s_{k,1}),g_{k,1}(s_{k,1}),\dots, g_{k,P}(s_{k,P})]^T $. In practice, to be more resilient against noise, we encode the resultant preprocessed vectors $g_k(\mathbf{s}_k)$ of length $P$ by the nested lattice codes to obtain $\mathbf{x}'_k =[x'_{k,1}, x'_{k,2},\dots, x'_{k,n}]^T  \in \Lambda_n \subset \mathbb{R}^{n\times 1}$  \cite{Gaspar2011} and denote $\mathbf{x}'[p] = [x'_{1,p}, x'_{2,p},\dots, x'_{K,p}]^T$, as shown in Fig. \ref{CoMACSystem01}. We employ a filter $\mathbf{B}_k$ on $\mathbf{x}'_k$ at each sensor node with the objective of minimizing sum mean-squared error of computed functions. Hence, each sensor node $k$ transmits $\mathbf{x}_k$ through NOMA channel, as shown in Fig. \ref{CoMACSystem01}. The main objective of the AP characterizes in decoding the received vector $\mathbf{y} =[y_{1}, y_{2},\dots, y_{n}]^T$ into $P$ desired functions (\ref{nomographicF}). In order to discuss our proposed optimization technique based on MMSE criterion, we first present a transmitter model over the NOMA channel, as discussed below. 
		\vspace{-0.4cm}
		
			\begin{center}
		\begin{figure*}
			\centering
			\includegraphics[width=5.5 in]{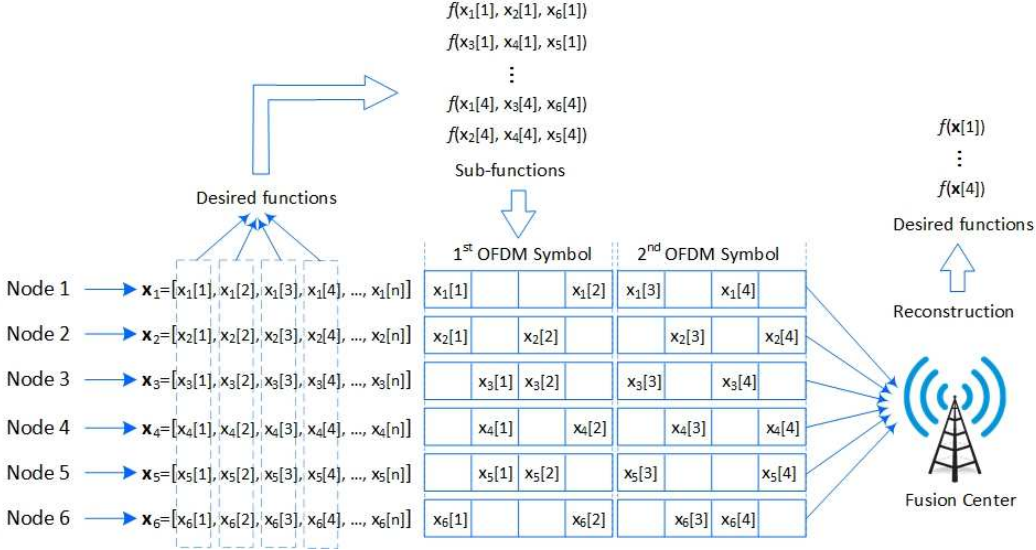}
			\centering \caption{Framework of wideband CoMAC.} \label{CoMAC01}
			\vspace{-0.2cm}
		\end{figure*}
	\end{center}
	
	\vspace{-0.4cm}
	
\subsection{NOMA Scheme}
We consider a wideband NOMA scheme with $N$ subcarrier over $T_s$ OFDM symbols for transmitting $\mathbf{x}_k$ at each sensor nodes. Due to fact that NOMA spreading waveforms are sparse only part of the $K$ sensor nodes participate in the computation at each subcarrier. Thus, the desired functions $f_p(\cdot)$ in (\ref{nomographicF}) that is composed of all $K$ sensor nodes can be broken down into subfunctions, as detailed in \cite{wu2020noma}. The subfunction is only part of the desired function, which is computed by a subset of $K$ sensor nodes.
Each subfunctions considers only $M$ chosen sensor nodes as a distinct subset of all $K$ nodes. Therefore, the desired function is split into $B = \frac{K}{M}$ parts. In each subcarrier, $L$ subfunctions are chosen such that $L = \frac{B}{D}$, where $D\in \mathbb{N}$. The desired functions are reconstructed by these subfunctions at the AP. Then, the $m$-th received OFDM symbol at AP can be formulated as \cite{wu2020noma}
\begin{equation}
\label{mainSystem}
\mathbf{Y}[m] = \sum_{l=1}^L \sum_{k =1}^K \mathbf{V}_k^l[m] \mathbf{X}_k^l[m] \mathbf{H}_k[m] +\mathbf{W}[m],
\end{equation}
\noindent where $m \in [1:T_s]$, $T_s = \frac{n}{N}$, $N$ is the number of subcarriers and $T_s$ is the number of OFDM symbols. The power allocation matrix of the $k$-th sensor node is denoted as $\mathbf{V}_k^l[m] = \mathsf{diag}\{ v_{k,1}^l[m], \dots, v_{k,N}^l[m]\}$, whose diagonal element is the power allocated to compute the $l$-th function at each subcarrier, $\mathbf{X}_k^l[m] = \mathsf{diag}\{ x_{k,1}^l[m], \dots, x_{k,N}^l[m]\}$ is the transmitted diagonal matrix of the $k$-th sensor node to compute the $l$-th function, a diagonal matrix $\mathbf{H}_k[m] = \mathsf{diag}\{ h_{k,1}[m], \dots, h_{k,N}[m]\}$ is the channel matrix in which the diagonal elements are the channel response of each subcarrier for node $k$ and the diagonal element of $\mathbf{W}[m]$ is identically and independently distributed (i.i.d.) complex Gaussian random noise. Due to linearity and diagonal matrix structure, we re-write (\ref{mainSystem}) as 
\begin{equation}
\label{mainSystema}
\mathbf{Y}[m] = \sum_{k =1}^K \sum_{l=1}^L  \mathbf{V}_k^l[m] \mathbf{X}_k^l[m] \mathbf{H}_k[m] +\mathbf{W}[m].
\end{equation}
 We define the combined matrix as follows:
\begin{equation}
\label{Xi}
\mathbf{X}_k[m] \eqdef \sum_{l=1}^L  \mathbf{V}_k^l[m] \mathbf{X}_k^l[m],
\end{equation}
\noindent where $\mathbf{X}_k[m] \!=\! \mathsf{diag}\{ x_{k,j_1}[m]v_{k,j_1}[m], \dots, x_{k,j_T}[m]v_{k,j_T}[m]\}$, $j_t \in \{1, 2, \dots N\}$, $t \in \{1,2,\dots, T\}$, $j_{t_a}\neq j_{t_b} $, and $T = LM$. The $T$ chosen sensor nodes corresponds to the nodes with the largest channel gains, e.g., $|h_{j_1}|\geq |h_{j_2}|\geq \dots \geq |h_{j_T}|$. We substitute (\ref{Xi}) into (\ref{mainSystema}) to obtain
\begin{equation}
\label{mainSystemb}
\mathbf{Y}[m] = \sum_{k =1}^K  \mathbf{X}_k[m] \mathbf{H}_k[m] +\mathbf{W}[m].
\end{equation}
Since $\mathbf{X}_k[m]$ and $\mathbf{H}_k[m]$ are diagonal matrices for $1 \leq k \leq K$, (\ref{mainSystemb}) can be expressed equivalently as
\begin{equation}
\label{mainSystemc}
\mathbf{Y}[m] = \sum_{k =1}^K   \mathbf{H}_k[m] \mathbf{X}_k[m] +\mathbf{W}[m].
\end{equation}
For ease of transmission-power control and without loss of generality, the $m$-th OFDM symbols are assumed to be normalized to have unit variance, i.e., $\mathbb{E} \{\mathbf{X}_k[m] \mathbf{X}_k[m]^H\} =\mathbf{I}_N$ for $\forall k$ and $\forall m$, where $\mathbf{I}_N$ denotes the identity matrix of size $N$ by $N$. Ideally, we would like to receive $\mathbf{X}[m]$ as one-to-one mapping expressed as\vspace{-0.2cm}
\begin{equation}
\label{Xm}
\mathbf{X}[m] = \sum_{k =1}^K   \mathbf{X}_k[m].
\end{equation}\vspace{-0.0cm}
The proposed optimization formulation is discussed in the next section. 

\vspace{-0.1cm}
	\subsection{MMSE Filtering}
We consider the joint optimization of transmit and receive filtering under the MMSE criterion with the transmission power constraints. Let $\mathbf{A}[m] \in \mathbb{C}^{N \times N}$ denote the receiver MMSE filtering matrix for $m$-th OFDM symbol at the AP and $\mathbf{B}_k[m] \in \mathbb{C}^{N \times N}$ the transmit MMSE filtering matrix at sensor node $k$ for $m$-th OFDM symbol. Let the $m$-th OFDM symbol combined matrix notation in (\ref{Xi}) be  defined now as $\mathbf{X}'_k[m]$, since the $k$-th sensor node will apply MMSE filtering on $\mathbf{x}'_k$ instead of the $\mathbf{x}_k$, as in (\ref{mainSystem}). Hence, the $m$-th OFDM symbol received at AP can be expressed as
 \begin{equation}
 \label{ReceivedFusion}
 \mathbf{\hat{X}}[m] = \mathbf{A}[m]^H\sum_{k =1}^K   \mathbf{H}_k[m] \mathbf{B}_k[m] \mathbf{X}'_k[m] +\mathbf{A}[m]^H\mathbf{W}[m].
 \end{equation} \vspace{-0.1cm}
 
 The distortion error between estimated $\mathbf{\hat{X}}[m]$ and $\mathbf{{X}}[m]$, which quantifies the over-the-air computation performance can be measured by MSE defined as follows:
  \begin{equation}
 \label{MSE}
 \mathsf{MSE}(\mathbf{\hat{X}}[m], \mathbf{{X}}[m])\! =\! \mathbb{E}\left\{\mathsf{tr}(\mathbf{\hat{X}}[m]\!-\! \mathbf{{X}}[m])(\mathbf{\hat{X}}[m]\!-\! \mathbf{{X}}[m])^H\right\},
 \end{equation}
 \noindent where $\mathsf{tr}(\cdot)$ denotes the sum of elements on the main diagonal of the square matrix. For the sake of simplicity, we will drop the $m$ notation from our formulations but readers should bear in mind that the formulation contains $m$, which refers to $m$-th OFDM symbol. Substituting (\ref{ReceivedFusion}) and (\ref{Xm}) into (\ref{MSE}), the MSE can be explicitly written as a function of the transmitter and receiver MMSE filtering as follows: \vspace{-0.0cm}
   \begin{eqnarray}
 \label{MSE_A}
 \mathsf{MSE}(\mathbf{{A}}, \{\mathbf{B}_k\})\! \!&=&\! \!\sum_{k =1}^K \mathsf{tr}(\mathbf{A}^H \mathbf{H}_k \mathbf{B}_k - \mathbf{I}  )(\mathbf{A}^H \mathbf{H}_k \mathbf{B}_k - \mathbf{I} )^H \nonumber \\
 &+& \sigma_n^2 \mathsf{tr}(\mathbf{A}^H \mathbf{A} ),
 \end{eqnarray}\vspace{-0.0cm}
 due to the fact that $\mathbb{E} \{\mathbf{X}_k \mathbf{X}_k^H\} =\mathbf{I}_N$. Our main objective in (\ref{MSE_A}) is to find the set of matrices $\mathbf{{A}}, \{\mathbf{B}_k\}$ such that MSE is minimized. Based on the widely known approach, we constrain the matrix $\mathbf{A}$ to be orthonormal matrix. Furthermore, under the MMSE criterion, a positive scaling factor $\eta$, called denoising factor, is included in $\mathbf{A}$ for regulating the tradeoff between noise reduction and transmission-power control. Define $\mathbf{A} = \sqrt{\eta} \mathbf{F}$ with $\mathbf{F}$ being a tall unitary matrix and thus $\mathbf{F}^H\mathbf{F} = \mathbf{I}_N$. Then given the MSE in (\ref{MSE_A}) the MMSE filtering problem can be formulated as \vspace{-0.0cm}
\begin{equation}
\begin{aligned}
(\textbf{P1}) \min_{\eta, \mathbf{{A}}, \{\mathbf{B}_k\}} \quad & \mathsf{MSE}(\mathbf{{A}}, \{\mathbf{B}_k\})\\
\textrm{s.t.} \quad & ||\mathbf{B}_k ||^2 \leq P_0 \: \: \forall i\\
&\mathbf{A}^H \mathbf{A} = \eta \mathbf{I}_N    \\
\end{aligned}
\end{equation}\vspace{-0.0cm}
 The solution to (\textbf{P1}) can be shown to be $\mathbf{A}^* = \sqrt{\eta^*} \mathbf{F}^*$, $\mathbf{B}_k^* = \mathbf{A}_k^H(\mathbf{A}_k \mathbf{A}_k^H)^{-1} \: \: \: \forall k$ and $\eta^* = \mathsf{max}_k \frac{1}{P_0}\mathsf{tr}(\mathbf{F}_k \mathbf{F}_k^H)^{-1}$, where $\mathbf{F}_k = (\mathbf{F}^*)^H\mathbf{H}_k$, $\mathbf{A}_k = (\mathbf{A}^*)^H\mathbf{H}_k$, and $\mathbf{F}^* = \mathbf{V}_G$\cite{Huang2018}. 
 
 Let the effective channel coefficients matrix defined by $\mathbf{G}$ as follows: \vspace{-0.1cm}
 \begin{equation} \vspace{-0.1cm}
 \mathbf{G} = \sum_{k =1}^K\lambda_{min}(\Sigma_k^2)\mathbf{U}_k\mathbf{U}_k^H,
 \label{Gmatrix}
 \end{equation} \vspace{-0.1cm}
 \noindent where $\mathbf{U}_k$ is the left matrix of singular value decomposition (SVD) of $\mathbf{H}_k$, namely $\mathbf{H}_k = \mathbf{U}_k \mathbf{\Sigma}_k \mathbf{V}_k^H$ and SVD of $\mathbf{G}$ is defined as $\mathbf{G}  = \mathbf{V}_G \mathbf{\Sigma}_G \mathbf{V}_G^H$.
 

 \begin{figure*}[ht!]
\centering
	\begin{subfigure}{0.30\textwidth}
		\centering
		\includegraphics[height=4.6cm]{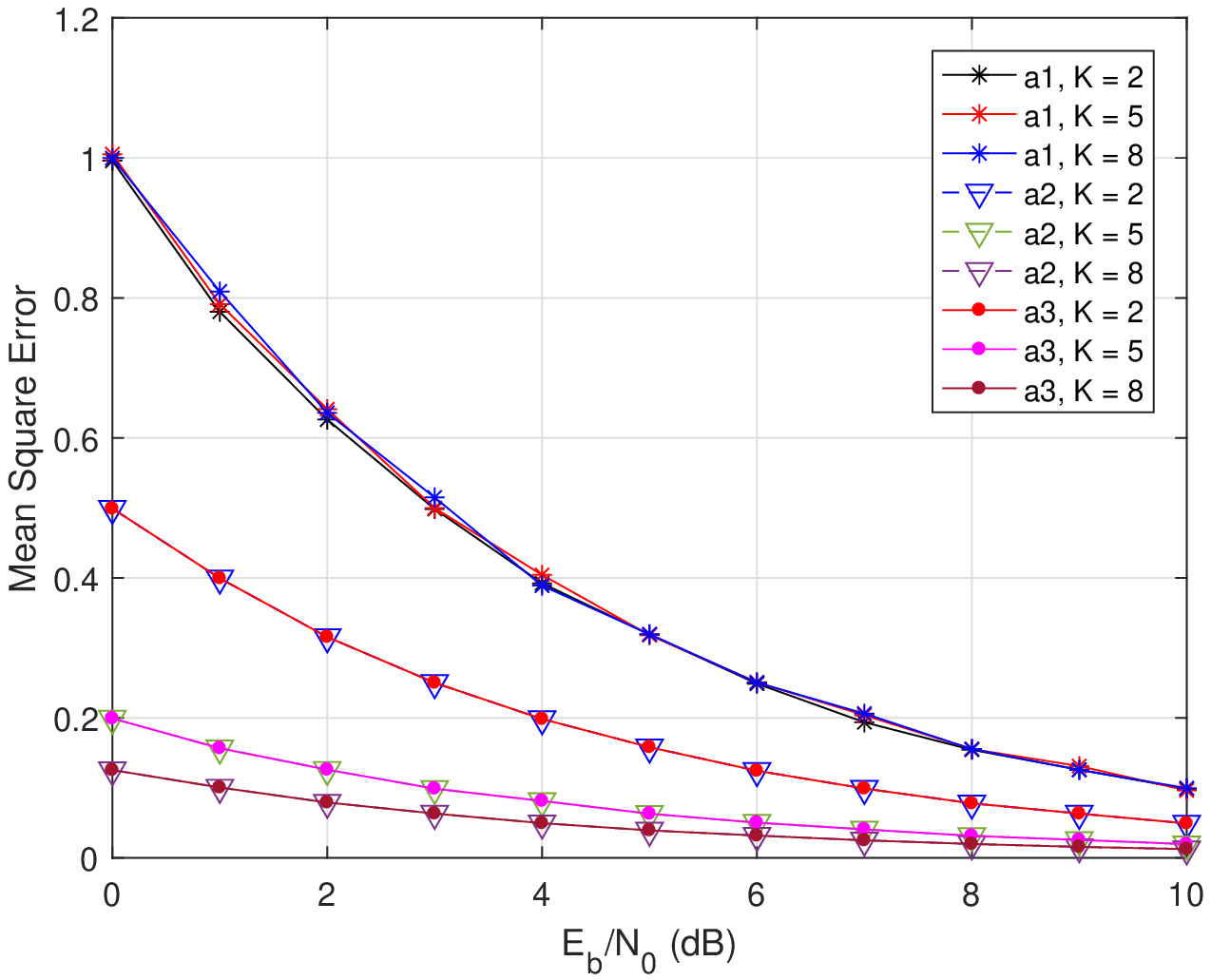}
		\caption{ }
 
	\end{subfigure}
	\begin{subfigure}{0.30\textwidth}
		\centering
		\includegraphics[height=4.6cm]{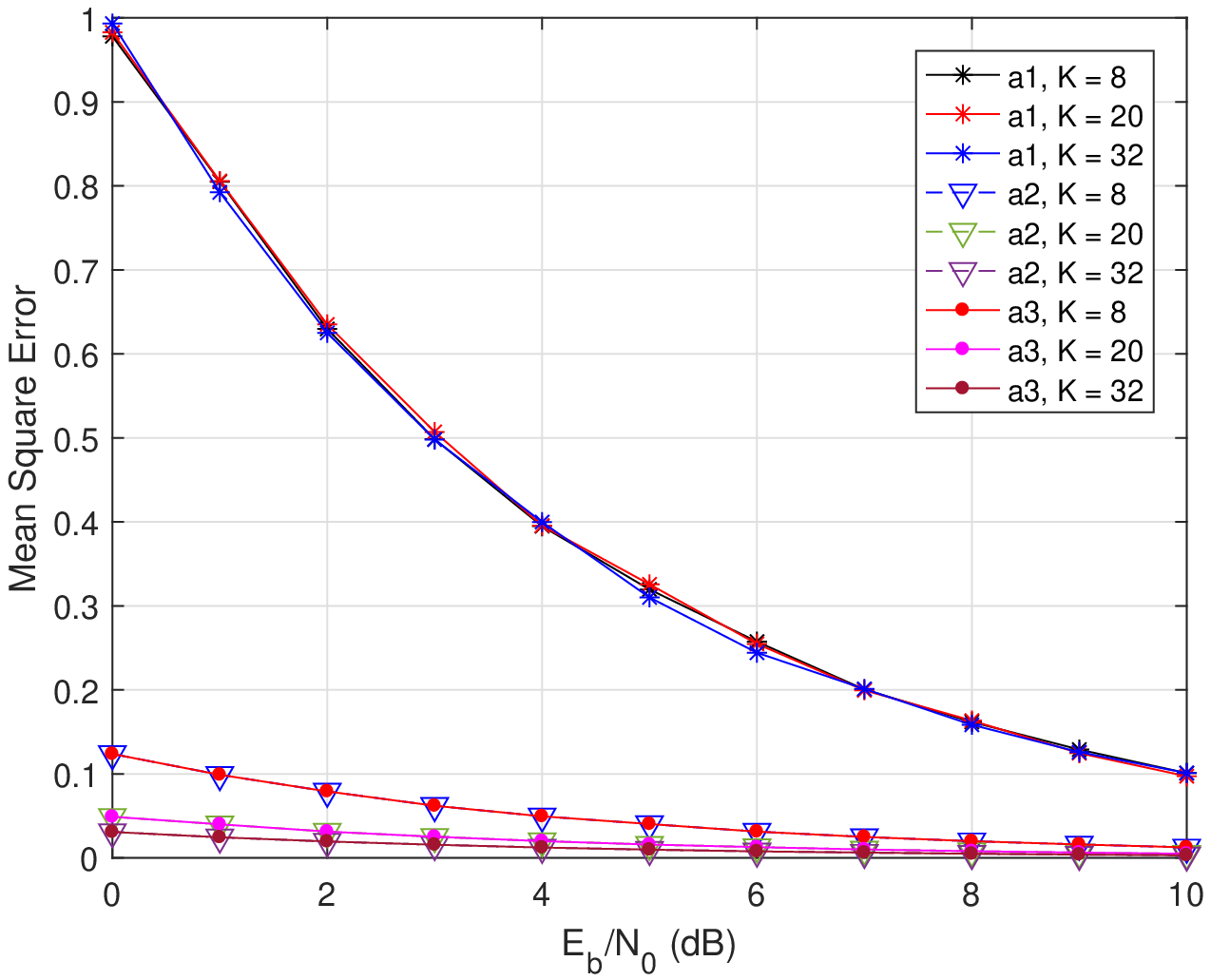}
		\caption{ }
	 
	\end{subfigure}
		\begin{subfigure}{0.30\textwidth}
		\centering
		\includegraphics[height=4.6cm]{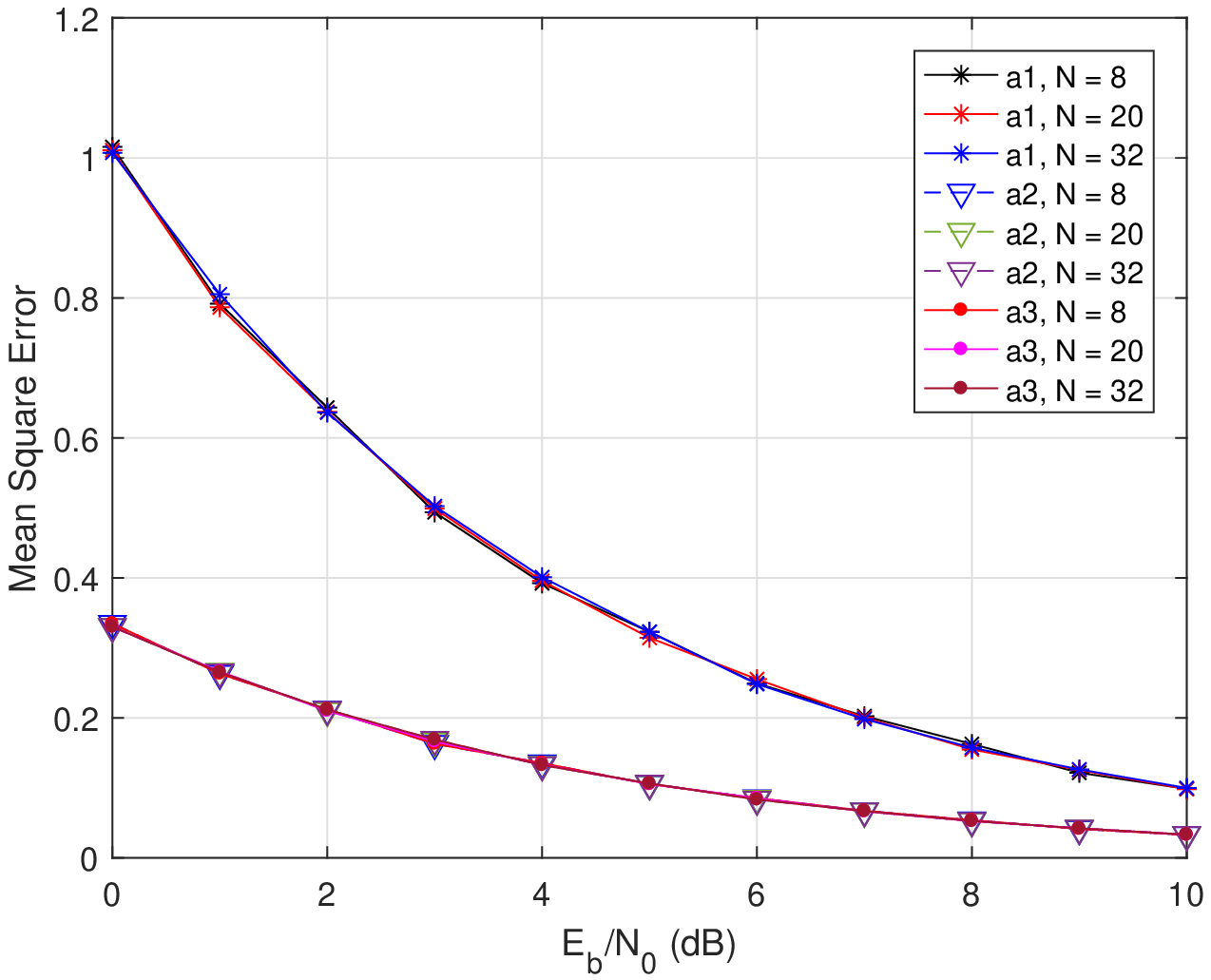}
		\caption{ }
		
	\end{subfigure}\vspace{-0.1cm}

	\caption{NOMA with  (a) $N = 6$ and $K = 2, 5, 8$, (b) $N = 12$ and $K = 8, 20, 32$, (c) $K = 3$ and $N = 8, 20, 32$}.
\label{3figs}
\end{figure*}

 \subsection{Eigenvector-Based Approach}
 \label{EigenVectorBased}
We consider an alternative solution of (\textbf{P1}) that is entirely based on the largest eigenvectors of the sum-channel matrix. Define the sum-channel matrix as \vspace{-0.1cm}
\begin{equation}
    \mathbf{H}_{s} = \sum_{k = 1}^K \mathbf{H}_k .
    \label{MatrixHs}
\end{equation}\vspace{-0.1cm}
The eigenvector-based solution can be achieved by $\mathbf{A}^* = \mathbf{Q}$, where $\mathbf{Q}$ is the eigenvector decomposition of $\mathbf{H}_s$ such that $\mathbf{H}_{s} = \mathbf{Q} \Lambda \mathbf{Q}^{-1}$. Note that $\mathbf{H}_s$ and $\mathbf{G}$ in (\ref{Gmatrix}) are diagonal matrices due to the fact that $\mathbf{H}_k$ is diagonal for $1\leq k \leq K$. Hence, it can be shown that the obtained solutions $\mathbf{A}^*$ are also diagonal indeed. More explicitly, it is an identity matrix, $\mathbf{I}_N$. We adopt the eigenvector-based approach as it is computationally less expensive to compute (\ref{MatrixHs}) than (\ref{Gmatrix}). 
\vspace{-0.1cm}
 \subsection{Channel Feedback Phase}
  The (\textbf{P1}) solution obtained requires perfect knowledge of global channel state coefficients $\{\mathbf{H}_k\}$ to be available at all $K$ sensor nodes. The proposed channel training and feedback mechanism, where we assume that the feedback observation at the AP can be noiseless is represented by  \vspace{-0.1cm}
 \begin{equation}
 \label{feedback}
 \mathbf{Z} =  \sum_{k =1}^K \mathbf{H}_k \mathbf{D}_k \: \: \in \mathbb{C}^{N \times N},
 \end{equation}\vspace{-0.1cm}
 \noindent where $\mathbf{D}_k\in \mathbb{C}^{N \times N}$ denotes the signal matrix transmitted by the sensor node $k$. Let $\mathbf{A}^*$ denoted the derived solution of (\textbf{P1}), $\tilde{f}(\cdot)$ and $\tilde{g}_k(\cdot)$ be the feedback counterparts of the pre- and post-processing operations of ${f}(\cdot)$ and ${g}_k(\cdot)$ for $1 \leq k \leq K$. One of the important design constraint is that the transmitted signal $\mathbf{D}_k$ in (\ref{feedback}) must be a function of $\mathbf{H}_k$ only, which we denote it as $\mathbf{D}_k = \tilde{g}_k(\mathbf{H}_k) $. Furthermore, the optimization problem can be formulated as\vspace{-0.1cm}
 \begin{equation} 
(\textbf{P2}) \: \: \:  \: \: \: \mathbf{A}^* = \tilde{f}(\sum_{k = 1}^K \mathbf{H}_k \tilde{g}_k(\mathbf{H}_k)),
 \end{equation}\vspace{-0.1cm}
 \noindent and the problem of feedback design reduces to the design of the functions $\tilde{f}(\cdot)$ and $\{\tilde{g}_i(\cdot)\}$. The solution is obtained when $\mathbf{Z} = \mathbf{G}$. Therefore, the feedback signal solution is obtained as $\mathbf{D}_k^* = \tilde{g}_k(\mathbf{H}_k) = \lambda_{min}(\mathbf{\Sigma}_k^2)\mathbf{V}_k \mathbf{\Sigma}_k^{-1} \mathbf{U}_k^H$ and feedback post-processing is $\mathbf{F}^* = \tilde{f}(\mathbf{Z}) = \mathbf{U}_Z$, where $\mathbf{U}_Z$ denotes the left eigenvectors of $\mathbf{Z}$. 
 \section{Simulation Results}
 \label{simulations}
 In this section, we evaluate the performance of proposed scheme via simulation studies. The MSE performance of (\ref{MSE}) are illustrated in Fig. \ref{3figs} for NOMA system. In Fig. \ref{3figs} (a), we set $N=6$ and vary $K=2$, $K=5$ and $K=8$, where $a1$, $a2$ and $a3$ denote average sum-channel based, eigenvector-based and effective channel based techniques, respectively. The MSE is a decreasing function of $E_b/N_o$.

 In addition, we observed that when $K$ increases MSE performance decreases further for the eigenvector-based and effective channel based techniques but not for the average sum-channel as portrayed in Fig. \ref{3figs} (b). Similar results are obtained for the fixed $N=12$ and varying $K=8$, $K=20$ and $K=32$ as shown in Fig. \ref{3figs} (a). In Fig. \ref{3figs} (c), we set $K=3$ and vary $N=8$, $N=20$ and $N=32$, respectively. We note that increasing $N$ does not effect the MSE performance unlike the case for increasing $K$ the MSE performance decreased for all solution methods as shown in Fig. \ref{3figs} (c). 
 
\begin{center}
	\begin{figure}[tb]
		\centering
		\includegraphics[width=3.2 in]{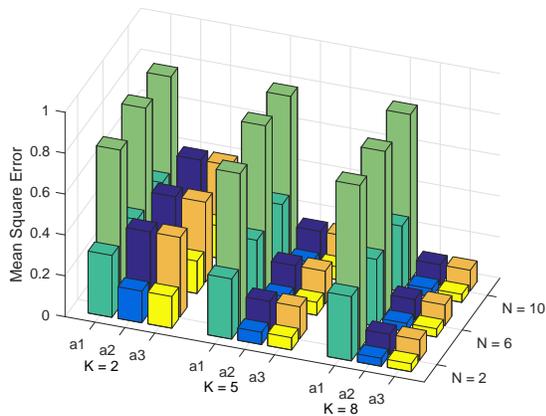}\vspace{-0.1cm}
		\centering \caption{NOMA with $K = 2, 5, 8$ and $N = 2,6,10$}. \label{3dBarK2_4N2_06_01}
		\vspace{-0.2cm} 
	\end{figure}
\end{center}

\begin{center}
	\begin{figure}[tb]
		\centering
		\includegraphics[width=3.2 in]{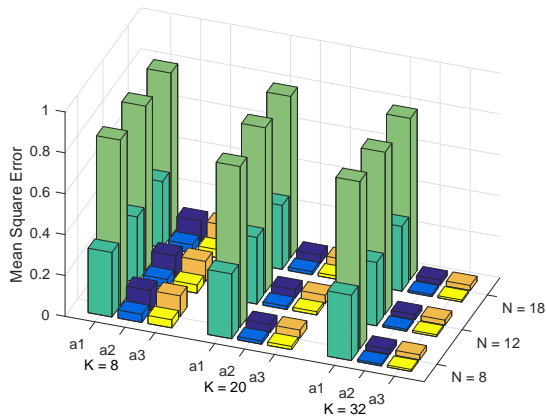}\vspace{-0.1cm}
		\centering \caption{NOMA with $K = 8, 20, 32$ and $N = 8,12,18$.} \label{3dBarK8_32N8_18_01} \vspace{-0.2cm}
	\end{figure}
\end{center}

 We further illustrate the effects on MSE performance by increasing the $K$ and $N$ values jointly in Figs. \ref{3dBarK2_4N2_06_01} and \ref{3dBarK8_32N8_18_01}, respectively. For each values of $K$ and $N$, we plot $6$ bar values where the front and back three values are for $a1$, $a2$ and $a3$ techniques evaluated at $E_b/N_o=1$ dB and $E_b/N_o=5$ dB, respectively. In all our simulation studies, we observe that the proposed eigenvector-based scheme outperforms the average sum-channel method, showing the effectiveness of new optimization based approach. Note the effective channel based and eigenvector-based methods have similar MSE performance, as discussed in Section \ref{EigenVectorBased}. Furthermore, the MSE performance gain of the proposed eigenvector-based scheme is more evident for larger values of $N$ and $K$, further confirming the effectiveness of the proposed solution for the multimodal sensing and dense networks. Our numerical results suggest that having a larger diversity gain is of great benefit, since it can provide a satisfactory MSE performance with reduced computational complexity.
    \vspace{-0.6cm}
 
\vspace{-0.4cm}

\vspace{-0.5cm}
\section{Complexity of Solution Methods}
    In this section, we will be focusing on the computation complexity of construction of filter $\mathbf{A}^*$.    The computational complexity of the average sum-channel based method, $a1$, is $\mathcal{O}(K^3)$, which involves the addition of $K$ matrices having the size of $K \times K$. The proposed eigenvector-based method, $a2$, in addition to $a1$, involves eigenvector decomposition process with the complexity of $\mathcal{O}(K^3)$, hence overall complexity is $\mathcal{O}(K^3+K^3) = \mathcal{O}(K^3)$. On the other hand, the complexity of effective channel based technique, $a3$, involves SVD for each matrix $\mathbf{H}_k$ with the complexity of $\mathcal{O}(K^3)$, $K \times K$ matrix multiplication and $K$ matrix addition that results in an overall complexity of $\mathcal{O}((K^3 + K^3)K) = \mathcal{O}(K^4)$. Since our matrices are diagonal in our formulations, it is straightforward to show that the complexity of sum-channel based method, $a1$, eigenvector-based method, $a2$, and effective channel based technique, $a3$, are $\mathcal{O}(K^2)$, $\mathcal{O}(K^2)$, and $\mathcal{O}(K^3)$, respectively, as shown in Table \ref{table:complexity}. 
		\vspace{-0.1cm}
		\begin{table}[tb]
	\caption{Computational Complexity Comparison}
	\centering 
	\begin{tabular}{l c c } 
		\hline\hline \rule{0pt}{3ex}  
		\bf{Algorithms} & \bf{Complexity} & \bf{Main procedures}  \\ [0.5ex]
		\hline \rule{0pt}{3ex}  
		$a1$& $\mathcal{O}(K^2)$ & multiplication, addition  \\[0.2ex]
		\:\:$a2$ &$\mathcal{O}(K^{2})$ & multiplication, addition   \\[0.2ex]
		\:\:$a3$ &$\mathcal{O}(K^3)$ & multiplication, addition   \\[0.2ex]
		\hline 
	\end{tabular}
	\label{table:complexity}
\end{table}

    In contrast to the $a3$ approach, the $a2$ approach has lower computation complexity although they both demonstrate similar MSE performance.
    \vspace{-0.0cm}
\section{Conclusion}
	\label{conclusion}
		In this paper, we developed a wideband non-orthogonal multiple-access based computation over multi-access channel (NOMA-CoMAC) framework. We formulated an optimization problem analytically in terms of the mean squared error (MSE), which is a prerequisite for the CoMAC, specifically in multimodal sensor networks.	We conceived an improved MSE minimization solution based on the eigenvector-based approach. 
		We demonstrated that our proposed eigenvector-based scheme achieves around $0.7$ lower at $E_b/N_o = 1$ dB in terms of MSE performance compared to average sum-channel based method. Moreover, MSE performance gain of our proposed solution increases for the larger values of $K$ by benefiting from the diversity gain. This, in return, suggests that our proposed scheme is eminently suitable for multimodal sensor networks.
		 In our future research, we will conceive NOMA-CoMAC for multimodal sensors for transmission over dispersive fading channels as well as possibility of incorporating network coding in the existing framework.

 \vspace{-0.4cm}

\bibliographystyle{IEEEtran}
\bibliography{CoMAC}

\end{document}